# A Computer-Aided Diagnosis System for Breast Pathology: A Deep Learning Approach with Model Interpretability from Pathological Perspective

Wei-Wen Hsu, Yongfang Wu, Chang Hao, Yu-Ling Hou, Xiang Gao,
Yun Shao, Xueli Zhang, Tao He, and Yanhong Tai*

*Abstract*— *Objective:* **We develop a computer-aided diagnosis (CAD) system using deep learning approaches for lesion detection and classification on whole-slide images (WSIs) with breast cancer. The deep features being distinguishing in classification from the convolutional neural networks (CNN) are demonstrated in this study to provide comprehensive interpretability for the proposed CAD system using pathological knowledge.** *Methods:* **In the experiment, a total of 186 slides of WSIs were collected and classified into three categories: Non-Carcinoma, Ductal Carcinoma in Situ (DCIS), and Invasive Ductal Carcinoma (IDC). Instead of conducting pixel-wise classification into three classes directly, we designed a hierarchical framework with the multi-view scheme that performs lesion detection for region proposal at higher magnification first and then conducts lesion classification at lower magnification for each detected lesion.** *Results:* **The slide-level accuracy rate for three-category classification reaches 90.8% (99/109) through 5-fold cross-validation and achieves 94.8% (73/77) on the testing set. The experimental results show that the morphological characteristics and co-occurrence properties learned by the deep learning models for lesion classification are accordant with the clinical rules in diagnosis.** *Conclusion:* **The pathological interpretability of the deep features not only enhances the reliability of the proposed CAD system to gain acceptance from medical specialists, but also facilitates the development of deep learning frameworks for various tasks in pathology.** *Significance:* **This paper presents a CAD system for pathological image analysis, which fills the clinical requirements and can be accepted by medical specialists with providing its interpretability from the pathological perspective.**

## I. INTRODUCTION

THE pathological examination has been the gold standard for diagnosis in cancer. It plays an important role since cancer diagnosis and staging help determine treatment options. However, the manual process of slide assessment is laborious and time-consuming. Wrong interpretations may occur due to fatigue or stress in specialists. Besides, the shortage of registered pathologists becomes a server problem in many countries. Consequently, the workload for pathologists has been increased and become unaffordable. Therefore, the computer-aided diagnosis (CAD) systems are developed to assist pathologists in slide assessment and work as a second opinion system to alleviate the workload of pathologists and avoid missing inspections.

Recently, the task performance in object recognition and image classification has significantly advanced as a result of the development of deep learning techniques [1]. Since 2012 [2], the framework of Deep Convolutional Neural Networks (DCNN) has shown its outstanding performance in many applications of computer vision. Because the DCNN framework is a representation learning approach that is suited for image analysis in digital pathology [3], it has been widely adopted in the development of CAD systems for tasks such as mitosis detection, lymphocyte detection and sub-type classification. And many studies have shown that the approaches using the features learned by the deep learning models, known as deep features, outperform the methods with the conventional hand-crafted features in histology image analysis [3-6].

In digital pathology, glass slides with tissue specimens were digitized by the whole-slide scanner at high resolution, becoming whole-slide images (WSIs) [7]. The analysis of WSIs is non-trivial because it involves a large amount of data (gigapixel level) to process and visualize [8]. In the preliminary stage, experiments [6, 9-12] were performed on open-access datasets from BreaKHis [13] or Bioimaging 2015 challenge [14] / BACH [4], in which the microscopy images for each class were cropped from the WSIs beforehand for training and testing instead of taking WSIs as the inputs directly. However, their approaches did not meet the clinical needs since cropping several regions of interests (ROIs) manually and sending them into the CAD system are almost infeasible in real clinical practice. To fill the clinical requirements, a fully automated diagnosis system that takes WSIs as the inputs and provides diagnostic assessment in both lesion-level and slide-level to assist pathologists in clinical practice is highly demanded.

For the CAD systems that process WSIs directly, Wang et al. [15] and Liu et al. [16] worked on the dataset from CAMELYON challenge [17] to detect breast cancer metastases on WSIs of lymph node sections. And the workable frameworks using deep learning approaches were proposed by Janowczyk and Madabhushi [3] for several detection tasks in pathology through WSIs. In the work by Bejnordi et al. [18], the clustering algorithm was adopted to detect ductal carcinoma in situ (DCIS) for WSIs with breast cancer. Even though the system performance was evaluated in both lesion-level and slide-level rigorously in their work, the developed system that simply detects lesions with DCIS may be inadequate for clinical use. The detection for invasive breast cancers may be more demanded from the clinical perspective since they are more server and tend to have a poorer prognosis.

On the other hand, Cruz-Roa et al. [19] applied the deep learning framework to detect invasive breast cancers in WSIs. The system performance was evaluated by the Dice Coefficient, which is a common assessment method in tasks of semantic segmentation (pixel-wise classification). Because it is hard to manually define precise and consistent contours for the invasive lesions, fair performance evaluation in lesion-level became almost infeasible. However, such an assessment of segmentation is more meaningful from the technical perspective than the clinical perspective.

From late 2017 to early 2018, the Grand Challenge on BreAst Cancer Histology images (BACH) [4] was launched, and the pixel-wise labeling in whole-slide breast histology images with three classes (benign, in situ carcinoma, and invasive carcinoma) was performed in the second task. The 3-category segmentation results can be very useful as a second opinion system for pathologists in clinical practice; nonetheless, the performance evaluation of segmentation in the challenge may be less meaningful for clinical purposes than the assessment in lesion-level and slide-level. Accordingly, to meet the clinical requirements, the performance of the proposed CAD system in both lesion-level and slide-level will be evaluated in this paper.

Besides, most developers of the state-of-the-art CAD systems with deep learning approaches treat deep learning models like a "black box" and only focus on the outcomes. As a result, the key features or the morphological characteristics learned by the deep learning models that contributed to the outcomes in lesion detection and classification have rarely been discussed, especially for breast cancer histopathology. A comprehensive mechanism of feature extraction in the deep learning framework for lesion classification using pathological explanations is focused in this study. Therefore, the reliability of the proposed CAD system can be enhanced, gaining acceptance from medical experts at the same time.

In this paper, we propose a deep learning framework that (1) integrates lesion detection and lesion classification using multi-view scheme and (2) offers the prediction accuracy in lesion-level and slide-level for performance evaluation to fill the clinical requirements. More importantly, the deep features learned by the deep learning models for lesion classification will be discussed, and the mechanism of the framework will be explained by the domain knowledge of pathology.

## II. MATERIALS AND METHODOLOGY

### A. Materials

In this study, a total of 186 H&E stained samples of breast tissues were collected from three categories: Non-Carcinoma, Ductal Carcinoma in Situ (DCIS), and Invasive Ductal Carcinoma (IDC). All samples were digitized to the format of WSIs by the scanner of ZEISS Axio Scan.Z1 at high resolution (x40). Lesion regions were labeled with 3 classes (Non-Carcinoma, DCIS, and IDC) by an experienced pathologist in 109 slides for cross-validation, and the other 77 cases were classified by the slide category for testing. All labeled regions were double-checked by the second registered pathologist, and the ground truth classification for each slide was determined by the diagnosis assisted with immunohistochemical results.

### B. The Development of CAD System

We aim to develop a fully automated CAD system that provides complementary and objective assessments to assist pathologists in diagnosis. To achieve that, a multi-view scheme was applied in our design that patches were sampled at different resolutions for analysis. Under higher magnification, the morphological characteristics of nuclei are focused for lesion detection. On the contrary, the spatial arrangement of cells or the co-occurrence properties of tissues is suited for observation at lower magnification. Accordingly, a hierarchical framework was designed, as shown in Fig. 1, for our CAD system to (1) perform lesion detection at high magnification for region proposal firstly and (2) classify each proposed region into Non-Carcinoma, DCIS, and IDC at intermediate to low magnification for lesion classification afterward.

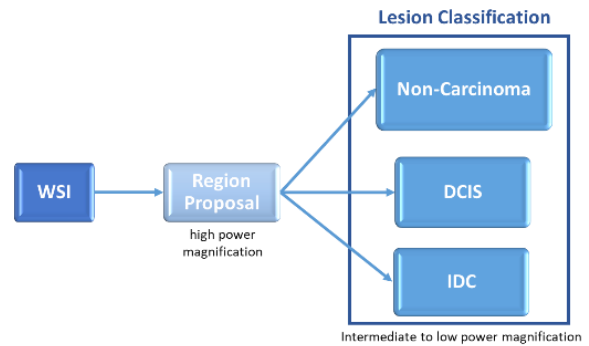

Fig. 1. The designed framework for the diagnosis of WSIs with breast cancer.

In the implementation, patches were sampled at different resolutions from the 109 slides with lesion regions labeled. For region proposal, a total of 130k patches were sampled at higher magnification from all 109 cases to fine-tune a pre-trained AlexNet-like model [20]. For lesion classification, 82k patches were sampled at intermediate to low magnification for training and testing through 5-fold cross-validation using the pre-trained model of ResNet50 [21]. Fig. 2 shows the training procedure in the deep learning framework for lesion classification.

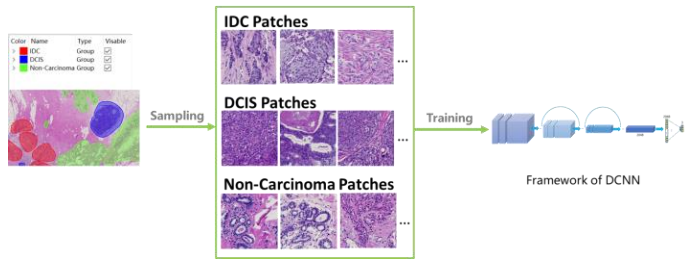

Fig. 2. The training procedure in the deep learning framework for lesion classification.

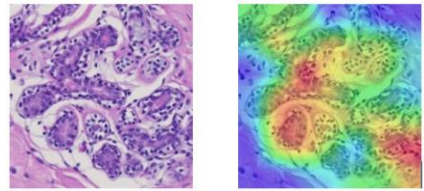

Fig. 3. An input patch (left) and its corresponding activation map overlaid on the input patch (right). The learned filter detects the distribution of lobules in the patch.

For slide assessment, the detection model was applied to perform lesion detection first. Then, the detected regions were filtered by the minimum size limit of 1 mm for region proposal since the case of microinvasive carcinoma was not considered in this study. The detection of microinvasion requires the framework with more precise localization algorithm rather than a patch-base method (patching). For each candidate region, several patches were randomly sampled from the region at lower magnification for lesion classification. The number of sampling patches depends on the area of the lesion. Afterward, the predictions of these sampling patches from the classification model determined which category the region belongs to using majority voting. Finally, the slide-wise classification can be achieved by analyzing the types of lesions that exist in the WSI.

### C. Performance Evaluation

The cross-validation set comprised 109 slides with lesion regions labeled. We computed the classification accuracy rates in patch-level, region-level, and slide-level through 5-fold cross-validation for performance analysis. Each WSI in the cross-validation set was unseen by its corresponding testing model for assessment, and all cases were executed in the testing procedure to derive the overall accuracy rate of classification. For the evaluation in region-level, the detected regions after size filtering were further classified into three categories (Non-Carcinoma, DCIS, and IDC) by two registered pathologists separately. If a region was labeled differently, the final decision was made after they reached a consensus.

To further validate the slide-wise performance of the proposed CAD system, all sampling patches from the cross-validation set (109 cases) were combined together for training. Then, the trained model was applied to the 77 cases in the testing set for slide assessment.

### D. Visualization of the Deep Features for Interpretability

To observe the semantic features that were learned by the deep learning model in classification, the input patches with high activation units were collected from the selected channel [22-25], and the corresponding activation maps were also generated to reflect the distribution of the matching patterns that the networks look for in the input images. Fig. 3 shows an example of an input patch with the highest activation unit in channel No.1134 and its corresponding activation map. From observations, the learned filter of no.1134 detects the distribution of lobules in the input patch, and it is distinctive to the category of Non-Carcinoma in lesion classification.

For feature analysis, the method of decision stump [26] was adopted to rank the top 100 discriminant features out of all 2048 features extracted by the model of ResNet50 for each category. The activations of all 2048 units from all sampling patches were computed and collected as a training dataset (#patches by 2048 features) for binary classification using decision stump. For each category in the classification, the classifier of decision stump derived an optimal threshold that led to the best classification results in every feature. If there strikes a high classification accuracy in a specific feature, it implies the feature is distinctive to the category so that the data within different classes can be well-separated. It infers what features are discriminative and contributory to classification. Furthermore, the trained weights of each feature for each class were also taken into consideration for analysis. The technique of Class Activation Mapping (CAM) [27] was applied to reflect the influential patterns in the input patch that are decisive in classification

## III. EXPERIMENTAL RESULTS AND DISCUSSION

### A. Performance of the CAD System

For lesion classification, the performance in patch-level, region-level, and slide-level through 5-fold cross-validation is provided in this study for evaluation and analysis, as shown in TABLE I. For 5-fold cross-validation, a set of 109 WSIs was divided into 5 datasets with slides evenly distributed in each category.

In patch-wise training and testing, a total of 82k patches were sampled from three different labeled regions (Non-Carcinoma, DCIS, and IDC). In 5-fold cross-validation, there were five datasets, and each dataset was treated as a testing set sequentially. If one of the datasets was assigned to be the testing set, the other four datasets were merged together to train a model. As a result, every sampling patch was blindly tested by a trained model. For binary classification, the models were used to predict whether there exists carcinoma in the input patch or not. For 3-category classification, the patches were categorized into Non-Carcinoma, DCIS, and IDC. In the experiment, the overall accuracy rate in patch-level reaches 89.3% for binary classification and 80.2% for 3-category classification.

In region-wise testing, approximately 2.5k regions were proposed from the phase of lesion detection after screening by the region size. All proposed regions were classified into 3 categories by two separate registered pathologists, and the agreement between them was 0.951 by Cohen's kappa coefficient. For those regions with different labels, the final ground truth was determined after coming to an agreement. The binary classification accuracy rate in region-level is 93.7%, and the 3-category classification accuracy rate is 91.1%. It is noteworthy that the region-level accuracy was largely improved compared with the accuracy in patch-level, which should be attributed to the design of the hierarchical framework and the scheme of majority voting in our CAD system. Instead of performing DCIS or IDC detection through WSIs directly, we designed the hierarchical framework that samples patches at different resolutions for lesion detection and classification for two main reasons: (1) The multi-view approach that lesions are observed under different power magnifications is highly accordant to the manual process of slide assessment in practice. The morphological properties are more apparent at higher magnification, while the architectural features are best seen under lower magnification. Accordingly, lesion detection and lesion classification were performed under different sampling views. (2) The scheme of majority voting in region-wise testing can achieve better error tolerance for the CAD system. Since not every sampling patch in testing can present the core properties of the lesion and the patch-wise predicting model is not perfect, applying majority voting for each proposed region can reduce the impacts of those misclassified patches effectively. Therefore, the error tolerance of the proposed system was improved by reducing the influences of misinterpretations. The lesions of lobular hyperplasia, Usual Ductal Hyperplasia (UDH), DCIS, and IDC that were recognized by the proposed system correctly in the lesion-wise classification are shown in Fig. 4. The results show the multi-view scheme and hierarchical framework can obtain better delineations of lesions and more consistent prediction results in classification.

TABLE I
SYSTEM PERFORMANCE IN PATCH-LEVEL, REGION-LEVEL, AND SLIDE-LEVEL THROUGH 5-FOLD CROSS-VALIDATION

| 5-fold cross-validation | Non-Carcinoma | Carcinoma | | Overall (109 cases) |
|---|---|---|---|---|
| | | DCIS | IDC | |
| **Patch-level** | 0.822 (20867/25394) | 0.925 (52804/57093) | | 0.893 (73671/82487) |
| | | 0.808 (13574/16808) | 0.787 (31687/40285) | 0.802 (66128/82487) |
| **Region-level** | 0.970 (1373/1416) | 0.893 (958/1073) | | 0.937 (2331/2489) |
| | | 0.847 (611/721) | 0.804 (283/352) | 0.911 (2267/2489) |
| **Slide-level** | 0.881 (37/42) | 0.970 (65/67) | | 0.936 (102/109) |
| | | 0.862 (25/29) | 0.974 (37/38) | 0.908 (99/109) |

Finally, the slide assessment can be accomplished by analyzing the lesion types that exist in each WSI. The slide-wise predicting accuracy rate is 93.6% for binary classification and 90.8% for 3-category classification through cross-validation. To further verify the robustness of the proposed CAD system for clinical practice, the total patches from the cross-validation set were used for training, and another 77 WSIs with confirmed slide labels were collected for testing. The prediction results are listed in TABLE II. The accuracy rates in slide-level are 97.4% and 94.8% for binary classification and 3-category classification respectively.

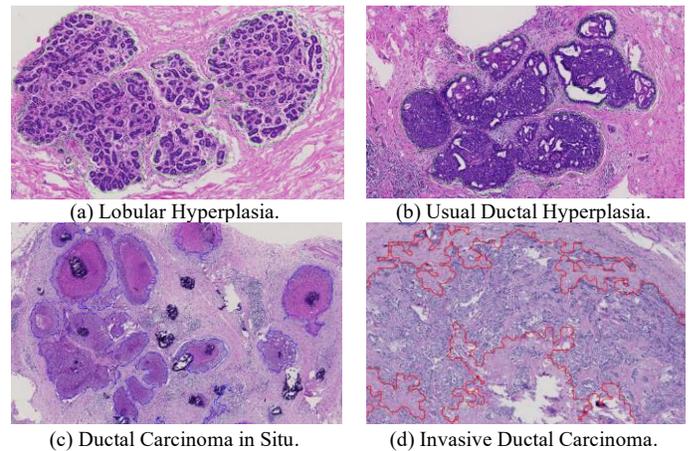

(a) Lobular Hyperplasia.  (b) Usual Ductal Hyperplasia.

(c) Ductal Carcinoma in Situ.  (d) Invasive Ductal Carcinoma.

Fig. 4. Correct interpretations from the proposed CAD system in the lesion-wise classification.

For all 186 cases in the experiment, there are four false negative cases that were misclassified by the system. Two of them are the cases with low-grade DCIS, and another two cases contain the lesions of low-grade IDC. Fig. 5 shows the examples of the misclassified lesions that caused false negatives in diagnosis. To tackle the issue of misinterpretations on cases with low-grade DCIS and low-grade IDC, more samples of these lesion types should be collected to enhance the training models regardless of their incidence rates. Also, the category of Atypical Ductal Hyperplasia (ADH) may need to be considered in lesion classification. Even though ADH shares many similarities with low-grade DCIS, many experts suggest it can be separated from low-grade DCIS by the quantitative criteria (2 mm in aggregate) [28, 29], and that can be done by the CAD system easily.

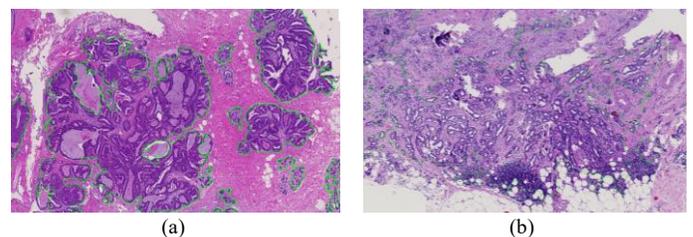

(a)  (b)

Fig. 5. False negative cases that contain the lesions of (a) low-grade DCIS and (b) low-grade IDC.

TABLE II
THE SLIDE-WISE PREDICTING PERFORMANCE ON THE TESTING SET

| Testing | Non-Carcinoma | Carcinoma | | Overall (77 cases) |
|---|---|---|---|---|
| | | DCIS | IDC | |
| Slide-level | 1 (34/34) | 0.953 (41/43) | | 0.974 (75/77) |
| | | 0.727 (8/11) | 0.969 (31/32) | 0.948 (73/77) |

The common misclassified lesions that caused false alarms are shown in Fig. 6. In Fig. 6-(a), invasive-growth-like patterns are present in the lesions of lobular cancerization (cancerization of lobules, COL) surrounded by lymphocytes and collagen that led to the misinterpretation. Because COL and IDC are morphologically different in detail, more patch samples from COL should be collected for training to capture the discriminative features. The area of cautery/fulguration artifacts [30, 31] is demonstrated in Fig. 6-(b). Since no training patches were sampled from these areas with cautery artifacts, these artifacts may not be recognized by the CAD system and become a potential cause of misinterpretation. Accordingly, the impacts of artifacts should also be considered in the development of CAD systems. In this case, those regions with cautery artifacts should have been excluded in the phase of region proposal by the detection model since the cautery artifacts contained no appreciable cellular architecture but heat-induced coagula [32].

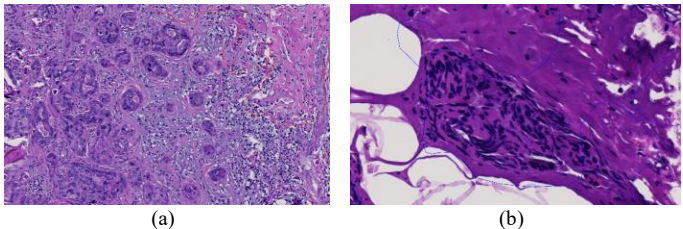

(a)                              (b)

Fig. 6. False alarms happened due to (a) the invasive-growth-like patterns in the lesions of lobular cancerization and (b) deformed cells and tissues from cautery artifacts.

### B. Visual Interpretability for Lesion Classification

Recently, deep learning frameworks have been widely adopted to perform high-stake tasks in many medical applications. However, without providing a robust interpretation for system's mechanism, a lack of validity of the system will become a problem. In this paper, the highly discriminant features for each category and their corresponding matching patterns are presented to provide the visual interpretability for the proposed CAD system.

In the pre-trained model of ResNet50, a total of 2048 features were used in classification. A patch with high activation on a unit implies a specific pattern being detected in the input image, and its corresponding activation map shows the pattern's spatial distribution. In the experiment, the most distinctive feature for the category of Non-Carcinoma took place on unit No.1134, denoted as feature No.1134 here. In the binary classification using decision stump, the feature No.1134 was used to distinguish the Non-Carcinoma patches from the patches with carcinoma lesions. Within deep feature No.1134, the activations between Non-Carcinoma and Carcinoma are well-separated, and its in-sample classification accuracy rate reaches 83.0% by the classifier of decision stump (simply using one feature), as shown in in Fig. 7.

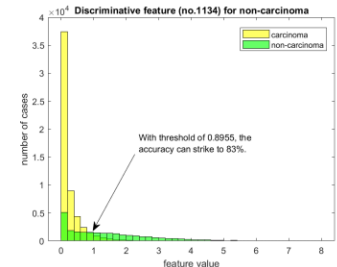

Fig. 7. Activations between Non-Carcinoma and Carcinoma are well-separated within feature No.1134.

The patches with high activation on unit No.1134 and the corresponding activation maps are shown in Fig. 8-(a). From observations, the feature No.1134 looks for lobules in the input patch. The learned weights of feature No.1134 for each category are listed in TABLE III. The positive weight $W_0$ implies that the detection of lobules is advantageous to classify the patch to the category of Non-Carcinoma. On the other hand, if the patch has a high activation on unit No.1134, it means that lobules were detected in the patch and its negative weights $W_1$ and $W_2$ tended to avoid the system from classifying the patch to the categories of DCIS and IDC. Except for lobules, the deep feature of wavy collagen fibers, as shown in Fig. 8-(b), and the morphological characteristic of benign epithelial hyperplasia, as shown in Fig. 8-(c), are also exclusive to the category of Non-Carcinoma. Similarly, their corresponding trained weights listed in TABLE III show positive $W_0$ and negative $W_1$, $W_2$ as the supportive features for the category of Non-Carcinoma.

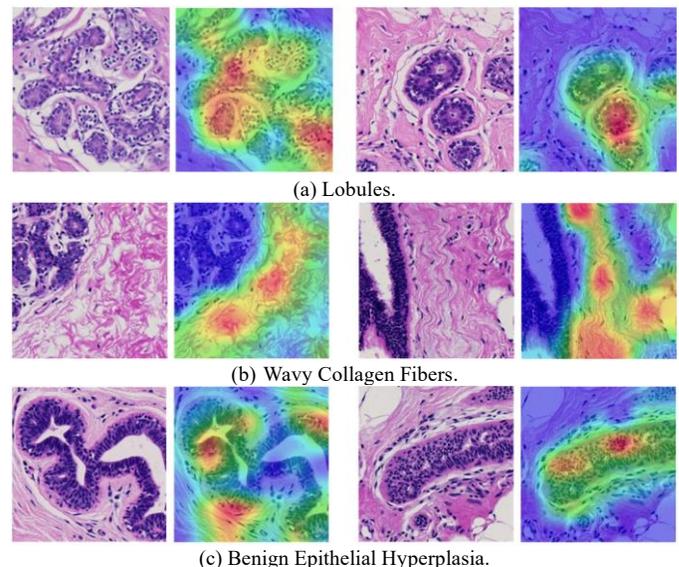

(a) Lobules.

(b) Wavy Collagen Fibers.

(c) Benign Epithelial Hyperplasia.

Fig. 8. The supportive features for the category of Non-Carcinoma.

TABLE III
FEATURE WEIGHTS FOR EACH CLASS LEARNED FROM THE
TRAINING DATA (POSITIVE $W_0$)

| Feature Number | $W_0$ (Non-Carcinoma) | $W_1$ (DCIS) | $W_2$ (IDC) | Matching Pattern |
|---|---|---|---|---|
| No. 1134 | **0.081** | -0.051 | -0.045 | lobules |
| No. 1833 | **0.053** | -0.018 | -0.04 | wavy collagen fibers |
| No. 685 | **0.048** | -0.037 | -0.014 | epithelial hyperplasia |

In the deep learning model, the final classification decision was made by all 2048 features with trained weights taken into consideration. Fig. 9 shows an example of an input patch from Non-Carcinoma using the CAM analysis to reflect the influential patterns that contribute to the classification results. As it is shown in Fig. 9, the feature of wavy collagen fibers can be very useful in classification especially when the lesions of benign epithelial hyperplasia partially appear in the sampling patch. In pathology, Tumor-Associated Collagen Signatures (TACS) are used to classify the distinctive patterns of collagen reorganization that occur during breast cancer progression. For normal and benign (TACS-1) cases, collagen is characterized to appear wavy and curly [45, 46]. The results show that the deep learning model focused on not only the morphological characteristics of cells but also the co-occurrence properties from interstitial portions around lesions, mimicking an experienced pathologist.

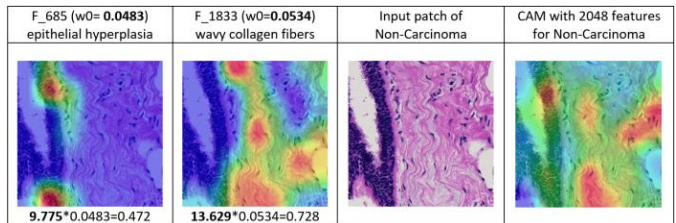

Fig. 9. The CAM analysis for the category of Non-Carcinoma.

For the category of DCIS, the most distinctive feature took place on unit No. 1815, and it detects the distribution of comedo-like necrosis in the sampling patches, as shown in Fig. 10-(a). With the classifier of decision stump, the in-sample classification accuracy rate achieves 84.3% in classifying DCSI patches or the others. Similarly, the learned weights for feature No.1815 in TABLE IV show that it is a discriminative feature for the category of DCIS (positive $W_1$ and negative $W_0$, $W_2$). In Fig. 10-(b), the cribriform growth pattern for the lesions of cribriform DCIS was captured by the deep learning model. Besides, the detection of the rounded configuration of lesions shown in Fig. 10-(c) can also contribute to the category of DCIS in classification. From observations, in general, the system recognizes the lesions of DCIS by checking if there exist cribriform growth patterns, intraluminal necrosis, or the solid nests with rounded configuration (different from the lesions of microinvasion or IDC). The approach of the deep learning model in recognizing lesions of DCIS agrees with the clinical rules in histopathology.

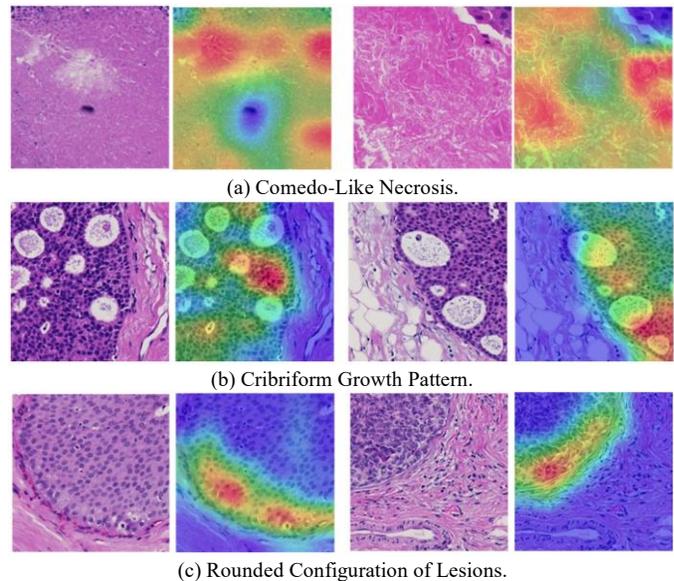

(a) Comedo-Like Necrosis.

(b) Cribriform Growth Pattern.

(c) Rounded Configuration of Lesions.

Fig. 10. The supportive features for the category of DCIS.

Different from the aforementioned features in supporting Non-Carcinoma that the positive weights only happened on $W_0$ for the category of Non-Carcinoma and negative weights for other classes, for some supportive features of DICS listed in TABLE IV, positive weights were not exclusive to $W_1$. Even though the features like cribriform patterns and rounded configuration are very useful in detecting DCIS lesions, the statistical results show that similar patterns could also be observed in the lesions of Non-Carcinoma. Fig. 11 shows the lesions of UDH can exhibit peripheral cribriform patterns and rounded configuration.

TABLE IV
FEATURE WEIGHTS FOR EACH CLASS LEARNED FROM THE
TRAINING DATA (POSITIVE $W_1$)

| Feature Number | $W_0$ (Non-Carcinoma) | $W_1$ (DCIS) | $W_2$ (IDC) | Matching Pattern |
|---|---|---|---|---|
| No. 1815 | -0.029 | **0.087** | -0.05 | necrosis |
| No. 1956 | **0.008** | **0.016** | -0.046 | cribriform pattern |
| No. 1402 | **0.006** | **0.034** | -0.021 | rounded configuration |
| No. 1819 | -0.0408 | **0.0487** | **0.0028** | solid nests |

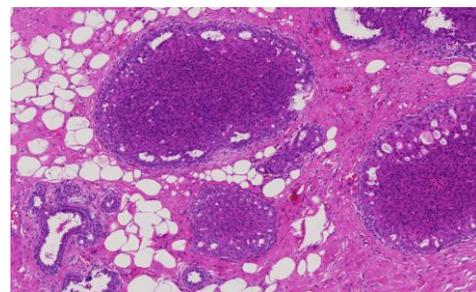

Fig. 11. The lesions of UDH with peripheral cribriform patterns and rounded configuration.

Even though the different categories of lesions may share the common deep features, the mechanism of the deep learning approaches can still distinguish different lesions and achieve lesion classification. In fact, the cribriform patterns in UDH and DCIS are morphologically different in detail, and it can be noticeable by the deep learning models. For example, the cribriform pattern in UDH reflects relatively weaker response than the one in DCIS within feature No. 1956, which constrains the likelihood of predicting a UDH lesion with cribriform patterns to the category of DCIS. On the other hand, for a cribriform lesion in DCIS, it is unlikely for the system to classify the lesion to Non-Carcinoma over DCIS since $W_1$ is larger than $W_0$ in feature No. 1956. More importantly, the deep learning models do not solely rely on one feature in decision making; instead, all features were taken into consideration in classification. The CAM analysis in Fig. 12 shows that the deep learning model recognizes the DCIS lesions by checking the co-existence of (a) solid nests and cribriform patterns or (b) solid nests and rounded configuration. If the solid nests were detected in the input patch, the negative weight $W_0$ of feature No. 1819 tended to avoid the system from classifying the patch to the category of Non-Carcinoma.

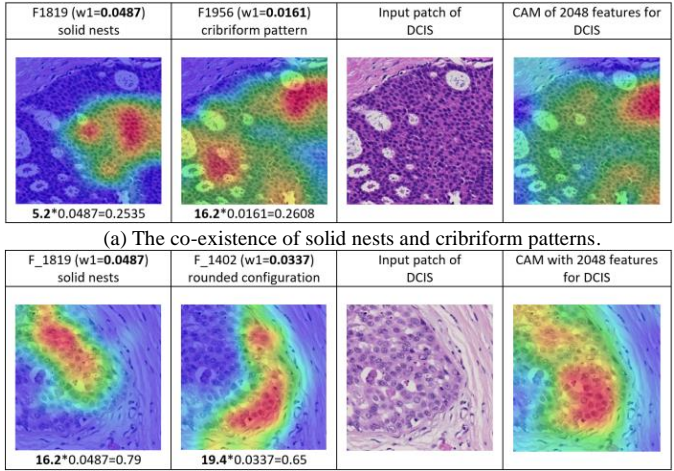

(a) The co-existence of solid nests and cribriform patterns.

(b) The co-existence of solid nests and rounded configuration.

Fig. 12. The CAM analysis for the category of DCIS.

The most distinctive feature for IDC is feature No.1261 in the experiment, as shown in Fig. 13-(a), which detects the cord growth patterns in the sampling patches. The in-sample classification accuracy rate is 83.6% in recognizing IDC patches using the cord growth feature by decision stump. In TABLE V, the weights of feature No.1261 also indicate the pattern is exclusive to the category of IDC (positive weight only happened on $W_2$). Fig. 13-(b) and Fig. 13-(c) illustrate the detected irregular clusters of tumor cells and solid nests with retraction spaces surrounded, which are also the distinctive features for IDC.

Fig. 14 shows that the co-occurrence property of solid nests and peritumoral retraction spaces is very crucial for the system to distinguish the lesions of IDC from DCIS. Since the weights of feature No.107 shows the solid patterns are supportive features for both DCIS and IDC, the existence of peritumoral retraction spaces becomes very discriminative in classification. In pathology, the study [33] has shown the peritumoral retraction artifact could be used as an inexpensive but effective invasive cancer marker. Such a phenomenon of tissue shrinkage of malignant tumor cells was also discovered by the deep learning models.

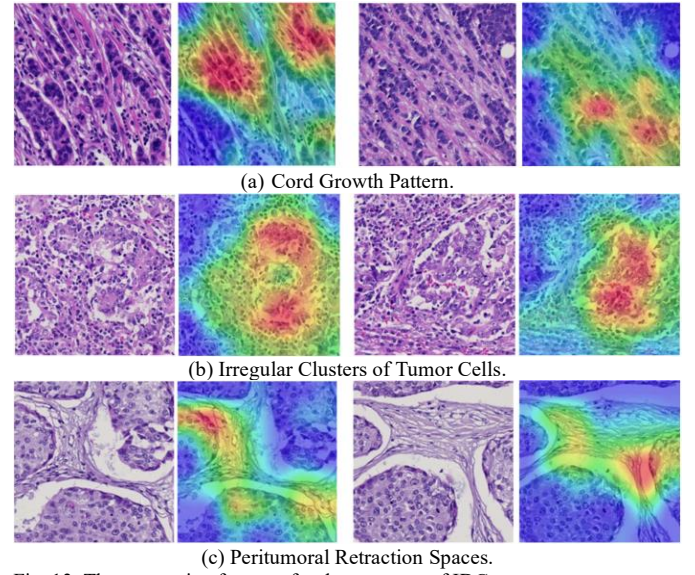

(a) Cord Growth Pattern.

(b) Irregular Clusters of Tumor Cells.

(c) Peritumoral Retraction Spaces.

Fig. 13. The supportive features for the category of IDC.

TABLE V
FEATURE WEIGHTS FOR EACH CLASS LEARNED FROM THE
TRAINING DATA (POSITIVE $W_2$)

| Feature Number | $W_0$ (Non-Carcinoma) | $W_1$ (DCIS) | $W_2$ (IDC) | Matching Pattern |
|---|---|---|---|---|
| No. 1261 | -0.013 | -0.02 | **0.031** | cord growth pattern |
| No. 1344 | -0.023 | -0.029 | **0.085** | irregular clusters of tumor cells |
| No. 1180 | -0.025 | -0.023 | **0.026** | peritumoral retraction spaces |
| No. 107 | -0.043 | **0.022** | 0.046 | solid nests |

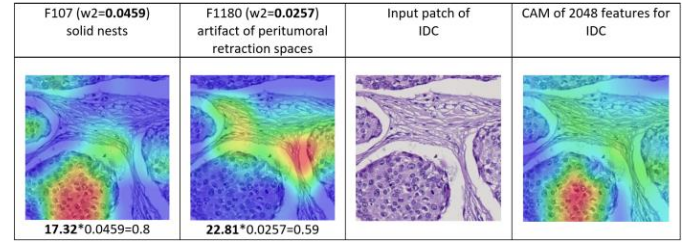

Fig. 14. The CAM analysis for the category of IDC.

The feature analysis has explained the mechanism of how a trained DCNN model recognizes different types of lesions, which follows the clinical rules in diagnosis. In lesion classification, the proposed CAD system not only focuses on the characteristics of cells' alignment but also learns the co-occurrence properties, such as the wavy collagen fibers in Non-Carcinoma, intraluminal necrosis in DCIS, and the peritumoral retraction spaces in IDC. The mechanism of the proposed system using deep learning approach is recognized by pathologists; therefore, the reliability and validity of the system can be established.

## IV. CONCLUSION

In this study, we propose a CAD system using deep learning approach for WSIs with breast cancer. The multi-view scheme and hierarchical framework were designed in the system to achieve better results in lesion-wise and slide-wise analysis. The proposed system can meet the clinical requirements, and its performance is analyzed in patch-level, region-level, and slide-level rigorously. Moreover, the mechanism of the deep learning approach in lesion classification is provided with the explanations from the pathological perspective. The results of visual interpretability are accordant with the clinical insights in lesion classification, which enhances the reliability and validity of our proposed system.

For future work, the false positive reduction can be done by decreasing the interference from the cautery artifacts in diagnosis. In addition, the proposed CAD system will be more versatile for the diagnosis of proliferative lesions if there are (1) further detection of microinvasive carcinoma for cases diagnosed as DCIS and (2) the category of ADH in lesion classification to avoid missed inspections for those high-risk precursor lesions and low-grade forms of DCIS.